\documentclass[showpacs,showkeys,floatfix,twocolumn]{revtex4}
\usepackage[cp1251]{inputenc}
\usepackage[english]{babel}
\usepackage[dvipdf]{color}
\usepackage{bm}
\usepackage{amsmath}

\begin{document}

\title{Asymptotic freedom at zero temperature in crystalline membranes}

\author{E.I.Kats, and V.V.Lebedev}

\affiliation{Landau Institute for Theoretical Physics, RAS, \\
142432, Chernogolovka, Moscow region, Russia, and \\
Moscow Institute of Physics and Technology, \\
141700, Dolgoprudny, Moscow region, Russia}

\begin{abstract}

We investigate effects of quantum (zero-temperature) long wavelength fluctuations of free
standing crystalline membranes, that are two-dimensional objects embedded into
three-dimensional space. The fluctuations produce logarithmic renormalization of
elasticity and bending moduli of the membranes. We find one-loop RG-equations to
demonstrate that the system is in the ``asymptotic freedom'' regime that is the quantum
fluctuations destabilize the flat membrane phase.

\end{abstract}

\pacs{68.55.J-, 68.35.Ct, 68.65.-k}

\maketitle

\section{Introduction}
\label{sec:intro}

Amazing electronic and mechanical properties of graphene \cite{NG04} and other
two-dimensional ($2d$) crystals \cite{NJ05} fuel continuously growing researches in this
area (see reviews \cite{NG09,VK10}, containing also numerous relevant references). This
progress also brings beautiful and new dimensions for researches not only restricted to
material science and its applications. It also shines some light and manifests some
analogies with basic phenomena of quantum field theory. As it is often stated in the
immense graphene literature, graphene might be served as a tool for realization or
visualization of high energy physics effects in a normal scale of solid state physics
laboratory. One could observe properties of charge carriers similar to ones of electrons
in quantum electrodynamics, e.g., the carrier mass renormalization \cite{NG09}).

Naturally, almost 10 years since graphene discovery and after huge number of
publications, one may wonder is there still something to be understood about graphene? It
is a purpose of our article to demonstrate that it is indeed so. Our investigation
concerns freely suspended graphene or other crystalline films. It turns out that
different $2d$ crystals can be stable and maintain macroscopic continuity and high
quality \cite{GG13}. Being freely suspended (substrate-free) such objects ($2d$ crystals
suspended in $3d$ space) can be considered as a new state of matter. In our paper, we
examine quantum (zero-temperature) long-scale fluctuations in such films and their
influence on macroscopic physical characteristics of the films.

Properties of the freely suspended crystalline films at finite temperatures are well
investigated, especially theoretically. Nelson and Peliti \cite{NP87} and Kantor and
Nelson \cite{KN87} were first suggesting that the anharmonic interaction produces a power
dependence on scale of the bending and elastic crystalline moduli of the films. The
prediction was confirmed later by systematic renormalization group (RG) calculations for
a four-dimensional crystalline membrane embedded into higher dimensional, $d>4$, space
\cite{AL88,PK88,AG89}. The approach becomes exact in the limit $d\to\infty$, see
\cite{DR92,GA09a,GA09b}. The power behavior of the moduli was checked by various scaling
and physical arguments and also atomistic simulations \cite{LK09,RF11,LK12}, and no
essential quantitative or qualitative contradictions to the theoretical predictions were
found.

Much less attention has been paid to quantum fluctuations in crystalline films. It is
partially because the quantum ($T=0$) non-linear effects lead to only logarithmic
corrections, hence generally much smaller than the power-law renormalization produced by
thermal fluctuations \cite{KA13}. However, small does not necessary mean irrelevant. In
this work we bring an attention of our readers to quantum fluctuations related to
vibrational modes. We claim that a freely suspended crystalline film manifests the
behavior, known as ``asymptotic freedom'' in the high energy physics. The asymptotic
freedom is a property that causes interactions between particles to become weaker as
energy increases (or distance between particles decreases), and at increasing space
scales the coupling constant grows. Note that the asymptotic freedom behavior is a
feature of quantum chromodynamics. This is one more illustration of a remarkable
peculiarity of the phenomenological Landau-like theoretical approach which is a powerful
tool for describing different systems irrespective to their microscopic nature.

Our paper is organized as follows. In Section \ref{sec:basic} we develop the low energy
(long wavelength) nonlinear theory of interacting vibrations in freely suspended $2d$
crystalline membrane. In Section \ref{sec:rg} one-loop renormalization group (RG)
equations for the elastic and bending moduli are derived and solved. We end the paper
with a discussion of possible physical consequences and interpretation of the results. We
relegate into Appendix technical steps of the RG-calculations.

 \section{Basic relations}
 \label{sec:basic}

We consider quantum fluctuations of a freely suspended crystalline film (membrane). The
membrane is treated as $2d$ infinitesimally thin film embedded into $3d$ space. Below we
assume that the membrane on average is perpendicular to the $Z$-axis. We also assume that
the membrane is symmetric that is both its sides are equivalent. Then in the main
approximation the long-scale energy (Landau functional) of the membrane can be written as
\cite{NP87,KN87}
 \begin{equation}
 {\cal F}=\int dx\, dy\, \left\{
 \mu w_{\alpha\beta}^2 +\frac{\lambda}{2}
 w_{\alpha\alpha}^2
 +\frac{\kappa}{2} (\nabla^2 h)^2 \right\}.
 \label{energy}
 \end{equation}
Here the coefficients $\mu,\lambda$ are $2d$ Lame (elastic) moduli, $\kappa$ is Helfrich
(bending) module of the membrane, $h$ is the membrane displacement along the
$Z$-direction, $w_{\alpha\beta}$ is the membrane in-plane distortion tensor, and the
subscripts $\alpha,\beta,\dots$ run over $x,y$. The distortion tensor can be written in
the main approximation as \cite{LL86}
 \begin{equation}
 2 w_{\alpha\beta}=
 \partial_\alpha u_\beta+\partial_\beta u_\alpha
 +\partial_\alpha h \partial_\beta h,
 \label{deften}
 \end{equation}
where $\bm u$ is the in-plane displacement vector of the membrane.

Analyzing quantum fluctuations one should investigate dynamics of the investigated
degrees of freedom. In the low-frequency limit the degrees of freedom (fields) to be
taken into account are $\bm u$ and $h$. The corresponding modes are the longitudinal and
the transverse sound modes with the acoustic dispersion laws $\omega=c_{l,t}k$ and also
the bending mode with the quadratic dispersion law $\omega=(\kappa/\rho) k^2$, where
$\rho$ is the $2d$ mass density of the membrane, $\omega$ is frequency and $\bm k$ is
(in-plane) wave vector. The sound velocities are expressed via the elasticity moduli as
 \begin{equation}
 c_l^2=\frac{2\mu+\lambda}{\rho}, \quad
 c_t^2=\frac{\mu}{\rho}.
 \label{defcc}
 \end{equation}
Note that optical vibrational modes may also exist in crystalline membranes (as, e.g., in
graphene). However, since the optical modes have a finite frequency they are irrelevant
for low-frequency effects we are investigating.

We are interested in correlation functions of the macroscopic fields $\bm u$ and $h$,
primarily in their pair correlation functions
 \begin{eqnarray}
 \langle u_\alpha(t,\bm r) u_\beta(0,0)\rangle
 =\int \frac{d\omega\ d^2k}{(2\pi)^3}
 e^{-i\omega t +i \bm k \bm r} F_{\alpha\beta}(\omega,\bm k),
 \label{corru} \\
 \langle h(t,\bm r) h(0,0)\rangle
 =\int \frac{d\omega\ d^2k}{(2\pi)^3}
 e^{-i\omega t +i \bm k \bm r}
 G(\omega, \bm k).
 \label{corrh}
 \end{eqnarray}
The above averages over quantum fluctuations (marked by angular brackets) can be
calculated as functional integrals over the fields $\bm u$ and $h$ with the weight
$\exp(iI/\hbar)$ where $I$ is the action
 \begin{eqnarray}
 I=\int dt\ d^2r\ \left\{
 \frac{\rho}{2} (\partial_t \bm u)^2
 +\frac{\rho}{2} (\partial_t h)^2 \right.
 \nonumber \\ \left.
 -\left[\mu w_{\alpha\beta}w_{\alpha\beta}
 +\frac{\lambda}{2} w_{\alpha\alpha}^2
 +\frac{\kappa}{2} (\nabla^2 h)^2\right]\right\},
 \label{action}
 \end{eqnarray}
reflecting dynamics of the membrane. The first term in the action (\ref{action}) is the
kinetic energy of the membrane whereas the second term there is the membrane potential
energy (\ref{energy}) (taken with the sign minus), both integrated over time. Note that
the expression (\ref{action}) is invariant under the transformation $h\to -h$ that
reflects the assumed symmetry of the membrane.

There are terms of the second, third and fourth order in the fields $\bm u$, $h$ in Eq.
(\ref{action}). The second-order term describes the non-interacting modes, whereas the
third-order and the fourth-order terms determine their interaction (coupling). The bare
correlation functions (\ref{corru},\ref{corrh}) determined by the quadratic (harmonic)
part of the action (\ref{action}) read as
 \begin{eqnarray}
 F_{\alpha\beta}(\omega,\bm k)=\frac{i\hbar}{\rho}
 \left[ \frac{1}{\omega^2 -c_l^2 k^2 +i0}
 \frac{k_\alpha k_\beta}{k^2} \right.
 \nonumber \\ \left.
 +\frac{1}{\omega^2 -c_t^2 k^2 +i0}
 \left(\delta_{\alpha\beta}
 -\frac{k_\alpha k_\beta}{k^2}\right) \right],
 \label{corrf} \\
 G(\omega, \bm k)= \frac{i\hbar}{\rho}
 \frac{1}{\omega^2- (\kappa/\rho) k^4 +i0}.
 \label{corrg}
 \end{eqnarray}
Here the term $+i 0$ implies the standard Feynman rule for the contour integration over
frequency near the poles of the Green functions. The positions of the poles correspond to
the dispersion laws of the acoustic modes $\omega = c_{l,t}k$ and of the bending mode
$\omega=(\kappa/\rho)^{1/2} k^2$.

Based on the interaction terms (of the third and fourth orders) in the action
(\ref{action}), one can develop a perturbation theory for the correlation functions of
the fields $\bm u$ and $h$. Fluctuation corrections, say, to the bare correlation
functions (\ref{corrf},\ref{corrg}) can be presented by Feynman diagrams with lines
corresponding to the bare correlation functions and vertices of the third and of the
fourth order determined by the interaction terms. One can check by direct calculations
that the perturbation theory produces logarithmic corrections to the parameters of the
bare correlation functions (\ref{corrf},\ref{corrg}). That is why below we use the
renormalization group (RG) technique to examine long-scale behavior of the correlation
functions of $\bm u$ and $h$.

 \section{Renormalization group equations}
 \label{sec:rg}

We use the renormalization group procedure in Wilson formulation, see Ref. \cite{WK74}.
The idea of the procedure is to split the fluctuating fields into fast (short-scale) and
slow (long-scale) parts and to integrate the distribution function (weight)
$\exp(iI/\hbar)$ over the fast component. As a result, we obtain a distribution function
for slow variables $\exp(iI'/\hbar)$ to be interpreted in terms of renormalized
parameters of the slow action $I'$.

Note that the distortion tensor (\ref{deften}) is invariant under the transformation
$\delta u_\alpha = - \theta _\alpha h$, $\delta h = \theta_x x+\theta_y y$ (where
$\theta_\alpha$ is an infinitesimally small rotation angle) that reflects the original
rotational invariance of the film. This symmetry leads to the conclusion, that the energy
and the ``potential'' part of the action have to be expressed in terms of the distortion
tensor (\ref{deften}). Consequently, the action $I'$ for the slow variables has the same
form (\ref{action}) but with renormalized parameters $\mu$, $\lambda$ and $\kappa$.

The renormalization of the Lame elastic moduli is determined by purely logarithmic
integrals. Based on the results presented in Appendix we end up with the following
quantum ($T=0$) one-loop RG-equations
 \begin{eqnarray}
 \frac{d\mu}{d\xi}=
 -\frac{\hbar}{32\pi \rho^{1/2}\kappa^{3/2}} \mu^2,
 \label{rgml1} \\
 \frac{d\lambda}{d\xi}=
 -\frac{\hbar}{32\pi \rho^{1/2}\kappa^{3/2}}
 (\mu^2+4\mu\lambda +2\lambda^2),
 \label{rgml2}
 \end{eqnarray}
where $\xi$ is logarithm of the scale. Renormalization of the bending modulus $\kappa$ is
more involved. We skip all rather tedious algebra (those readers who are interested in
mathematical details of the calculations can find all essential steps in Appendix), and
present the final result, the one-loop RG equation for the bending modulus $\kappa$
 \begin{equation}
 \frac{d\kappa}{d\xi}
 = -\frac{\hbar}{8\pi \rho^{1/2}\kappa^{1/2}}
 \frac{3\mu^2 + 3\mu\lambda}{2\mu+\lambda}.
 \label{corr14}
 \end{equation}
Thus bending oscillations become softer due to the quantum fluctuations (in contrast,
thermal fluctuations lead to hardening the oscillations).

One finds from Eqs. (\ref{rgml1},\ref{rgml2}) the one-loop RG-equation for the ratio of
the Lame moduli
 \begin{equation}
 \frac{d}{d\xi}\frac{\lambda}{\mu}=
 -\frac{\hbar \mu}{32\pi \rho^{1/2}\kappa^{3/2}}
 \left(1+3\frac{\lambda}{\mu} +2 \frac{\lambda^2}{\mu^2}\right).
 \label{rgrg2}
 \end{equation}
As it follows from the equation, there are two fix points of the ratio, $\lambda =-\mu/2$
and $\lambda=-\mu$. The fix point $\lambda = -\mu$ (corresponding to zero bulk modulus)
is unstable and we stay with the only stable fix point $\lambda =-\mu/2$. It is worth to
compare this finding with the known result for renormalization of Lame moduli by thermal
fluctuations \cite{AL88,PK88,AG89,DR92,GA09a,GA09b,LK09,RF11,LK12}. In the latter case
the RG-equations (formulated for a four-dimensional membrane) possess four different fix
points (including $\lambda = - \mu /2$) but the only stable fix point is $\lambda =- \mu
/3$. Since the stable fix points are not identical for the two cases, one should not
expect a sort of continuous matching of classical and quantum results.

Further we assume that the system (membrane) is in the state characterized by the ratio
near the fix point $\lambda =-\mu/2$. Substituting $\lambda =-\mu/2$ into Eq.
(\ref{corr14}) one obtains
 \begin{equation*}
 \frac{d\kappa}{d\xi}
 =-\frac{\hbar \mu}{8\pi\rho^{1/2}\kappa^{1/2}}.
 \end{equation*}
Then one finds from Eq. (\ref{rgml1})
 \begin{equation}
 \frac{dg}{d\xi}= g^2, \qquad
 g=\frac{5\hbar\mu}{32\pi \rho^{1/2}\kappa^{3/2}},
 \label{coupling}
 \end{equation}
where we introduced the dimensionless coupling constant $g$. The above RG-equations are
correct provided $g\ll 1$. We see that the coupling constant increases as the scale
grows. Therefore for large enough scales the system passes to the strong coupling regime
corresponding to strong fluctuations of the membrane shape.

Expanding the right-hand side of the equation (\ref{rgrg2}) near the fix point $\lambda
=-\mu/2$, we find
 \begin{equation*}
 \frac{d}{dg}\left(\frac{\lambda}{\mu}+\frac{1}{2}\right)
 =-\frac{1}{5g}\left(\frac{\lambda}{\mu}+\frac{1}{2}\right).
 \end{equation*}
Therefore the ratio $\lambda/\mu$ tends to $1/2$ as the coupling constant grows. It
justifies our approach. However, the corresponding law is $\lambda/\mu+1/2 \propto
g^{-1/5}$, that is the system approaches the fix point not too fast. Therefore, at
analyzing concrete experimental data, probably, it is worth to consider $\mu$ and
$\lambda$, as independent variables.

We conclude from Eq. (\ref{coupling}) that the coupling constant characterizing the
quantum fluctuations of the crystalline membranes, increases as the scale grows:
 \begin{equation*}
 g=g_0/(1-g_0 \xi).
 \end{equation*}
By other words, we encounter the ``asymptotic freedom'' behavior like in quantum
chromodynamics. This scenario means that unlike thermal fluctuations, stabilizing the
flat state of the crystalline membrane (hardening the classical bending rigidity),
quantum fluctuations yield to a rough (crumple) membrane state due to softening of
quantum bending fluctuations.

 \section{Conclusion}
 \label{sec:con}

What we found looks a bit counterintuitive: at $T=0$ crystalline membranes turn out to be
more rough at large scales in contrast to finite temperatures. Although there is no
theorem claiming that quantum fluctuations can be obtained as a sort of interpolation
with $T \to 0$ from classical thermal fluctuations, our finding may shake some of the
arguments used in analysis of thermal fluctuations by the very courages matching results
obtained for four dimensional membranes embedded in infinite dimensional space
\cite{DR92,GA09a,GA09b}.

A natural question appears concerning experimentally observable consequences of the
qualitatively surprising but quantitatively rather modest (logarithmic) renormalization
of the elastic moduli. For the graphene monoatomic films all physical parameters are
known \cite{NG09,VK10}. Namely, $\mu \simeq 9\, eV/\AA ^2$, $\lambda \simeq 2\, eV/\AA
^2$, $\kappa \simeq 0.7\, eV$, and 2$d$ density is $\rho \simeq 7.6 \cdot 10^{-8}\,
g/cm^2$. Combining everything together and stretching experimental uncertainty, our
estimations give $g \simeq 1/20$. The physical reason for the small value of $g$ is quite
transparent. A characteristic energy related to the elastic moduli ($\lambda$, $\mu$ and
$\kappa$) in the quantum limit is determined by the electronic mass $m$. On the other
hand, the membrane vibrations are related to Debye energy, that is are determined by the
atomic mass $M$. One can see that in fact $g \sim (m/M)^{1/3}$ (apart of numeric
factors).

Of course it is not very realistic to overcome the small bare coupling constant in
graphene by a large logarithmic factor. However, the situation is not completely
hopeless. We can remind the situation with smectic liquid crystals, where logarithmic
divergence of the layer displacements requires astronomic scales for its direct
observation but could be routinely observed by the power-law tails in X-ray scattering of
standard laboratory samples \cite{JO03}. Similarly, one can think, say, about
measurements of the simultaneous two-point correlation function $\langle\nabla h_1 \nabla
h_2\rangle$. The bare correlation function is proportional to $\delta (\bm r)$, where
$\bm r=\bm r_1-\bm r_2$ is the separation between the points. However, the logarithmic
renormalization produces corrections to the correlation function proportional to $r^{-2}$
(with some logarithmic factor). The small coefficient $g$ in front of this correction
makes its observation problematic but not impossible.

One more direction to think about observation of our findings is to get membranes with
larger values of the bare coupling constant $g$. One can try to use, e.g., freely
suspended crystalline smectic films with anomalously small bending elastic modulus
$\kappa$. Such behavior is expected in the systems undergoing transition into the
membrane ripple phase or into the so-called smectic $\tilde A$ or smectic $\tilde C$
structures with one-dimensional layer modulations. If the period of this modulation $p$
is larger than the molecular size $a$, then the bare bending modulus acquires a small
prefactor $(a/p)^2$. This factor for membrane ripple phase and for some modulated
smectics can be as large as $10^2$ (see, e.g., the survey \cite{BE91}, the papers
\cite{SV86,EG89}, and the more recent discussion of modulated structures in non-chiral
smectics \cite{VH13}).

We did not touch the electron-phonon coupling (or, more precisely, coupling to the
low-frequency modes we are investigating). The reason is that the Dirac electronic
degrees of freedom, despite their softness, do not yield to any additional long
wavelength renormalization of quantum vibrations of the crystalline membrane. Indeed, at
the Dirac point the electron-phonon coupling has the form of the so-called deformational
potential \cite{MO08,GK12}. The feed-back influence of this deformation potential into
the long wavelength vibrational part of the Hamiltonian is irrelevant, and therefore does
not change our main conclusion concerning asymptotic freedom behavior of quantum
vibrational fluctuations in free standing crystalline films.

We hope that our work will motivate further experimental and theoretical studies along
this line -- quantum fluctuations in freely suspended crystalline membranes.

\acknowledgements

We would like to thank M.I.Katsnelson for helpful communications. Our work was supported
by the RFBR through grants 13-02-00120, 13-02-01460 and programs of Russian Ministry of
Education and Science.

\appendix{}

 \section{}
 \label{sec:app}

Here we develop the RG-procedure starting from splitting the fields $\bm u$ and $h$ into
the fast and slow parts: $\bm u \to \bm u' + \tilde{\bm u}$, $h \to h' +\tilde h$, where
prime designates slow fields and tilde designates fast fields. One can assume that the
fast fields $\tilde{\bm u}$ and $\tilde h$ are sums of spacial harmonics with wave
vectors $\bm q$ in the interval $\Lambda' < q < \Lambda$, where $\Lambda$ is the
ultraviolet cutoff, and $\Lambda'$ is a separation wave vector of slow and fast
variables. Next, one should calculate a ``slow'' action $I'(\bm u',h')$ in accordance
with the definition
 \begin{equation}
 \exp(i I'/\hbar)=
 \int {\cal D}\tilde u\,
 {\cal D}\tilde h\,
 \exp(i I/\hbar),
 \label{step}
 \end{equation}
where functional integration over fast fields is implied.

The rotational symmetry discussed in the main body of the paper guarantees that the
``slow'' action $I'(\bm u',h')$ is determined by the same expression (\ref{action}) but
with renormalized parameters. Therefore one can calculate corrections only to the
harmonic (quadratic) part of the action
 \begin{eqnarray}
 I^{(2)}=\int dt\ d^2r\ \left\{
 \frac{\rho}{2} (\partial_t \bm u)^2
 +\frac{\rho}{2} (\partial_t h)^2 \right.
 \nonumber \\ \left.
 -\left[\mu (\partial_\alpha u_\beta)^2
 +\frac{\lambda}{2} (\partial_\alpha u_\alpha)^2
 +\frac{\kappa}{2} (\nabla^2 h)^2\right]\right\}.
 \label{harmon}
 \end{eqnarray}
Actually, there appear logarithmic corrections to the moduli $\mu$, $\lambda$, $\kappa$
whereas logarithmic corrections to the mass density $\rho$ are absent.

If the coupling constant $g$ (\ref{coupling}) is small then the renormalization of the
moduli $\mu$, $\lambda$, $\kappa$ can be calculated by a loop expansion. To find
corrections to the harmonic action (\ref{harmon}) in the main one-loop approximation it
is enough to use the third-order contribution to the action (\ref{action})
 \begin{eqnarray}
 I^{(3)}=-\int dt\, d^2r
 \left[\mu \partial_\alpha u_\beta \partial_\alpha h \partial_\beta h
 +\frac{\lambda}{2} \nabla \bm u (\nabla h)^2\right].
 \label{action3}
 \end{eqnarray}
Substituting here $\bm u = \bm u'+\tilde{\bm u}$, $h=h'+\tilde h$ we find the following
second-order in the fast fields interaction term in the action
 \begin{eqnarray}
 I_\mathrm{int}=-\int dt\, d^2r
 \left[\mu \partial_\alpha u'_\beta
 \partial_\alpha \tilde h \partial_\beta \tilde h
 +\frac{\lambda}{2} \nabla \bm u' (\nabla \tilde h)^2\right]
 \nonumber \\
 -\int dt\, d^2r
 \left[\mu \partial_\alpha \tilde u_\beta \partial_\alpha \tilde h \partial_\beta h'
 +\mu \partial_\alpha \tilde u_\beta \partial_\alpha h' \partial_\beta \tilde h \right.
 \nonumber \\ \left.
 +\lambda \nabla \tilde{\bm u} \nabla h' \nabla\tilde h \right],
 \label{actionin}
 \end{eqnarray}
needed to calculate the one-loop contribution to the harmonic action (\ref{harmon}).

Let us first consider the one-loop correction to the slow harmonic action produced by the
first term in Eq. (\ref{actionin})
 \begin{eqnarray}
 \Delta_1 I^{(2)}= \frac{i}{2\hbar} \int dt\ d^2 r\
 \left(\mu \partial_\alpha u'_\beta
 +\frac{\lambda}{2}\nabla\cdot \bm u' \delta_{\alpha\beta}\right)
 \nonumber \\
 \left(\mu \partial_\mu u'_\nu
 +\frac{\lambda}{2}\nabla\cdot \bm u' \delta_{\mu\nu}\right)
 \nonumber \\
 \int dt_1\ d^2 r_1\
 \langle \partial_\alpha \tilde h \partial_\beta \tilde h
 \partial_\mu \tilde h_1 \partial_\nu \tilde h_1 \rangle_0,
 \label{action4}
 \end{eqnarray}
that determines renormalization of the Lame coefficients $\mu$ and $\lambda$. Here the
subscript $0$ marks correlation functions found by averaging with the harmonic action
(\ref{harmon}). First we have to calculate the integral entering the correction
(\ref{action4}). It is simpler to calculate it in the Fourier representation
 \begin{eqnarray}
 \int dt_1\ d^2 r_1\
 \langle \partial_\alpha \tilde h \partial_\beta \tilde h
 \partial_\mu \tilde h_1 \partial_\nu \tilde h_1 \rangle_0
 \nonumber \\
 = - \frac{2\hbar^2}{\rho^2}
 \int \frac{d\omega\ d^2q}{(2\pi)^3}
 \frac{q_\alpha q_\beta q_\mu q_\nu}{[\omega^2- (\kappa/\rho) q^4 +i0]^2}
 \nonumber \\
 =-\frac{i \hbar^2}{32 \pi \rho^{1/2}\kappa^{3/2}}
 \ln\frac{\Lambda}{\Lambda'}
 (\delta_{\alpha\beta}\delta_{\mu\nu}
 +\delta_{\alpha\mu}\delta_{\beta\nu}
 +\delta_{\alpha\nu}\delta_{\beta\mu}).
 \nonumber
 \end{eqnarray}
Substituting the result into Eq. (\ref{action4}) one finds corrections to $\mu$ and
$\lambda$ leading to RG-equations (\ref{rgml1},\ref{rgml2}) presented in the main body of
the paper.

Now we pass to calculation of the one-loop renormalization of $\kappa$ determined by the
second term in Eq. (\ref{actionin}). The corresponding correction to the slow harmonic
action is
 \begin{eqnarray}
 \Delta_2 I^{(2)}= \frac{i}{2\hbar} \int dt_1\ dt_2\ d^2 r_1\ d^2 r_2\
 \partial_\alpha h'_1 \partial_\mu h'_2
 \nonumber \\
 \langle \partial_\beta \tilde h_1 \partial_\nu \tilde h_2 \rangle_0
 \langle(\mu \partial_\alpha \tilde u_{1\beta}
 +\mu \partial_\beta \tilde u_{1\alpha}
 \nonumber \\
 +\lambda \nabla\cdot \bm \tilde u_1 \delta_{\alpha\beta})
 (\mu \partial_\mu \tilde u_{2\nu}
 +\mu \partial_\nu \tilde u_{2\mu}
 +\lambda \nabla\cdot \tilde{\bm u}_2 \delta_{\mu \nu}) \rangle_0.
 \label{action5}
 \end{eqnarray}
The principal technical problem here is that the main contribution related to fast
variables is ultraviolet. Therefore one has to extract the logarithmic term on the top of
the ultraviolet contribution. To be confident about results of this rather involved
calculations we do it both in real space and in Fourier space.

\subsection{Real space calculations}

A characteristic time $t=t_1-t_2$ in Eq. (\ref{action5}) is determined by the $\langle
\tilde u \tilde u \rangle$ correlation time and is, consequently, much less than one of
the $\langle \tilde h \tilde h \rangle$ correlation time. Therefore one can substitute
there the simultaneous $\langle \tilde h \tilde h \rangle$ correlation function. Changing
variables as $\tau=(t_1+t_2)/2$, $\bm r= \bm r_1-\bm r_2$, $\bm R=(\bm r_1+\bm r_2)/2$
and expanding $\nabla h$ over $\bm r$, one finds
 \begin{eqnarray}
 \Delta_2 I^{(2)}= -\frac{i}{4\hbar} \int d\tau\ d^2 R\
 \partial_\alpha \partial_\gamma h
 \partial_\mu \partial_\delta h
 \nonumber \\
 \int d^2 r\ r_\gamma r_\delta
 \langle \partial_\beta \tilde h(0,\bm r_1)
 \partial_\nu \tilde h(0,\bm r_2) \rangle_0
 \nonumber \\ \int dt \
 \langle(\mu \partial_\alpha \tilde u_{1\beta}
 +\mu \partial_\beta \tilde u_{1\alpha}
 +\lambda \nabla\cdot \tilde {\bm u}_1 \delta_{\alpha\beta})
 \nonumber \\
 \langle(\mu \partial_\mu \tilde u_{2\nu}
 +\mu \partial_\nu \tilde u_{2\mu}
 +\lambda \nabla\cdot \tilde {\bm u}_2
 \delta_{\mu \nu}) \rangle_0,
 \label{action6}
 \end{eqnarray}
where $h=h(\tau,\bm R)$.

The correlation functions entering (\ref{action6}) can be extracted from
 \begin{eqnarray}
 \langle \tilde h(0,\bm r) \tilde h(0,0)\rangle_0
 =\frac{i\hbar}{\rho}
 \int \frac{d\omega\ d^2k}{(2\pi)^3}
 \frac{\exp(i \bm k \bm r)}
 {\omega^2 - (\kappa/\rho) k^4 +i0}
 \nonumber \\
 =\frac{\hbar}{2\rho^{1/2}\kappa^{1/2}}
 \int \frac{d^2 k}{(2\pi)^2}
 \frac{\exp(i\bm k \bm r)}{k^2}
 =\frac{\hbar}{4\pi\rho^{1/2}\kappa^{1/2}} \ln(L/r),
 \nonumber
 \end{eqnarray}
and
 \begin{eqnarray}
 \int dt\ \langle \tilde u_\alpha(t,\bm r) \tilde u_\beta(0,0)\rangle_0
 = -\frac{i\hbar}{\rho}
 \int \frac{d^2 k}{(2\pi)^2}\exp(i\bm k \bm r)
 \nonumber \\
 \left[\frac{k_\alpha k_\beta}{k^2}
 \frac{1}{c_l^2 k^2}
 +\left(\delta_{\alpha\beta}-\frac{k_\alpha k_\beta}{k^2}\right)
 \frac{1}{c_t^2 k^2} \right]
 \nonumber \\
 =-\frac{i\hbar}{4\pi (2\mu+\lambda)}\left\{
 \left[\ln(L/r)+1/2\right]\delta_{\alpha\beta}
 -\frac{r_\alpha r_\beta}{r^2}\right\}
 \nonumber \\
 -\frac{i\hbar}{4\pi \mu}\left\{
 \left[\ln(L/r)-1/2\right]\delta_{\alpha\beta}
 +\frac{r_\alpha r_\beta}{r^2}\right\}.
 \nonumber
 \end{eqnarray}
Taking space derivatives one finds
 \begin{eqnarray}
 \langle \partial_\beta \tilde h(0,\bm r_1)
 \partial_\nu \tilde h(0,\bm r_2) \rangle_0
 =\frac{\hbar}{4\pi\rho^{1/2}\kappa^{1/2}r^2}
 \left(\delta_{\beta \nu}-2 \frac{r_\beta r_\nu}{r^2}\right),
 \nonumber
 \end{eqnarray}
and
 \begin{eqnarray}
 \int dt\ \langle \partial_\mu \tilde u_\alpha(t,\bm r)
 \partial_\nu \tilde u_\beta(0,0)\rangle_0
 \nonumber \\
 =\frac{i\hbar}{4\pi (2\mu+\lambda) r^2}
 \left\{ -\left(
 \delta_{\mu\nu}\delta_{\alpha \beta}
 +\delta_{\mu\alpha}\delta_{\nu \beta}
 +\delta_{\mu\beta}\delta_{\nu\alpha}
 \right) \right.
 \nonumber \\
 +2\left(
 \delta_{\alpha \beta}\frac{r_\mu r_\nu}{r^2}
 +\delta_{\alpha \mu}\frac{r_\beta r_\nu}{r^2}
 +\delta_{\alpha \nu}\frac{r_\mu r_\beta}{r^2}
 +\delta_{\mu \beta}\frac{r_\alpha r_\nu}{r^2}
 \right.
 \nonumber \\ \left. \left.
 +\delta_{\nu \beta}\frac{r_\alpha r_\mu}{r^2}
 +\delta_{\mu \nu}\frac{r_\alpha r_\beta}{r^2} \right)
 - 8 \frac{r_\alpha r_\beta r_\mu r_\nu}{r^4} \right\}
 \nonumber \\
 +\frac{i\hbar}{4\pi\mu r^2}
 \left\{ -\left(
 \delta_{\mu\nu}\delta_{\alpha \beta}
 -\delta_{\mu\alpha}\delta_{\nu \beta}
 -\delta_{\mu\beta}\delta_{\nu\alpha}
 \right) \right.
 \nonumber \\
 +2\left(
 \delta_{\alpha \beta}\frac{r_\mu r_\nu}{r^2}
 -\delta_{\alpha \mu}\frac{r_\beta r_\nu}{r^2}
 -\delta_{\alpha \nu}\frac{r_\mu r_\beta}{r^2}
 -\delta_{\mu \beta}\frac{r_\alpha r_\nu}{r^2}
 \right.
 \nonumber \\ \left. \left.
 -\delta_{\nu \beta}\frac{r_\alpha r_\mu}{r^2}
 -\delta_{\mu \nu}\frac{r_\alpha r_\beta}{r^2} \right)
 + 8 \frac{r_\alpha r_\beta r_\mu r_\nu}{r^4} \right\}.
 \nonumber
 \end{eqnarray}

Substituting the above expressions into Eq. (\ref{action6}) one obtains
 \begin{eqnarray}
 \Delta_2 I^{(2)}= \frac{\hbar}{32\pi \rho^{1/2}\kappa^{1/2}}
 \int d\tau\ d^2 R\ \partial_\alpha\partial_\gamma h
 \partial_\mu \partial_\delta h
 \nonumber \\ 
 \int d^2r\ \frac{r_\gamma r_\delta}{r^2}
 \frac{1}{2\pi  r^2}
 \left[ \frac{16 \mu^2}{2\mu+\lambda}
 \left(-\delta_{\alpha \mu}+\frac{r_\alpha r_\mu}{r^2}\right) \right.
 \nonumber \\ \left.
 -\frac{8\mu\lambda}{2\mu+\lambda} \delta_{\alpha \mu}
 - 4\mu \delta_{\alpha \mu}\right]
 \nonumber \\
 =-\frac{\hbar}{16\pi \rho^{1/2}\kappa^{1/2}}
 \int \frac{d\tau\ d^2 R}{2\pi} (\nabla^2 h)^2 \ln\frac{\Lambda}{\Lambda'}
 \nonumber \\
 \left[\frac{\mu^2}{2\mu+\lambda}+\frac{2\mu\lambda}{2\mu+\lambda}
 +\mu\right].
 \label{corr13}
 \end{eqnarray}
The contribution (\ref{corr13}) to the effective action gives a correction to $\kappa$
that leads to the RG-equation (\ref{corr14}).

 \subsection{Fourier-space calculations}

In the subsection we start from the same correction to the action (\ref{action5}) that is
rewritten in Fourier-representation as
 \begin{eqnarray}
 \Delta_2 I^{(2)}= \frac{i}{2\hbar}
 \int \frac{d\omega\ d^2 k}{(2\pi)^3}
 k_\alpha k_\mu h(\omega,\bm k) h(-\omega,-\bm k)
 \nonumber \\
 \int \frac{d\nu\ d^2 q}{(2\pi)^3}
 (q_\beta+k_\beta)(q_\nu+k_\nu)
 G(\omega+\nu,\bm k+\bm q)
 \nonumber \\
 (\mu q_\alpha \delta_{\beta\kappa}
 +\mu q_\beta \delta_{\alpha\kappa}
 +\lambda q_\kappa \delta_{\alpha\beta})
 \nonumber \\
 (\mu q_\mu \delta_{\nu\lambda}+\mu q_\nu \delta_{\mu\lambda}
 +\lambda q_\lambda \delta_{\mu \nu})
 F_{\kappa\lambda}(\nu,\bm q).
 \label{four1}
 \end{eqnarray}
Since the bending mode is slower than the acoustic ones, we may substitute $F(\nu,\bm
q)\to F(0,\bm q)$ in Eq. (\ref{four1}). Integrating then over $\nu$, one obtains
 \begin{eqnarray}
 \Delta_2 I^{(2)}= \frac{i}{4\rho^{1/2}\kappa^{1/2}}
 \int \frac{d\omega\ d^2 k}{(2\pi)^3}
 k_\alpha k_\mu h(\omega,\bm k) h(-\omega,-\bm k)
 \nonumber \\
 \int \frac{d^2 q}{(2\pi)^2}
 (q_\beta+k_\beta)(q_\nu+k_\nu)(\bm q+\bm k)^{-2}
 \nonumber \\
 (\mu q_\alpha \delta_{\beta\kappa}
 +\mu q_\beta \delta_{\alpha\kappa}
 +\lambda q_\kappa \delta_{\alpha\beta})
 \nonumber \\
 (\mu q_\mu \delta_{\nu\lambda}+\mu q_\nu \delta_{\mu\lambda}
 +\lambda q_\lambda \delta_{\mu \nu})
 F_{\kappa\lambda}(0,\bm q).
 \nonumber
 \end{eqnarray}
Here $\bm k$ is the wave vector of the slow variables whereas $\bm q$ is the wave vector
of fast variables, therefore $q\gg k$.

In the main order in $k/q$ we obtain an ultraviolet integral for $\Delta_2 I^{(2)}$. To
extract the renormalization of $\kappa$ one has to expand the above expression for
$\Delta_2 I^{(2)}$ up to the second order in $\bm k$ to obtain
 \begin{eqnarray}
 \Delta_2 I^{(2)}= \frac{\hbar}{4\rho^{1/2}\kappa^{1/2}}
 \int \frac{d\omega\ d^2 k}{(2\pi)^3}
 \nonumber \\
 k_\alpha k_\mu h(\omega,\bm k) h(-\omega,-\bm k)
 \int \frac{d^2 q}{(2\pi)^2}
 \left[\frac{k_\beta k_\nu}{q^2} \right.
 \nonumber \\ \left.
 -2 \frac{q_\beta k_\nu+k_\beta q_\nu}{q^2}
 \frac{\bm q \bm k}{q^2}
 +4 \frac{q_\beta q_\nu}{q^2}
 \frac{(\bm q \bm k)^2}{q^4}
 -\frac{q_\beta q_\nu}{q^2}
 \frac{k^2}{q^2}\right]
 \nonumber \\
 (\mu q_\alpha \delta_{\beta\kappa}
 +\mu q_\beta \delta_{\alpha\kappa}
 +\lambda q_\kappa \delta_{\alpha\beta})
 \nonumber \\
 (\mu q_\mu \delta_{\nu\lambda}+\mu q_\nu \delta_{\mu\lambda}
 +\lambda q_\lambda \delta_{\mu \nu})
 \nonumber \\
 \frac{1}{q^2}\left[
 \frac{1}{2\mu+\lambda}\frac{q_\kappa q_\lambda}{q^2}
 +\frac{1}{\mu}\left(\delta_{\kappa\lambda}
 -\frac{q_\kappa q_\lambda}{q^2}\right)\right].
 \label{action22}
 \end{eqnarray}
Below, we separately calculate contributions to $\Delta_2 I^{(2)}$ related to the
longitudinal factor $q_\kappa q_\lambda/q^2$ and to the isotropic factor
$\delta_{\kappa\lambda}$ in the last line of Eq. (\ref{action22}). The ``longitudinal''
contribution to the slow action is
 \begin{eqnarray}
 \Delta_{21}I^{(2)}= \frac{\hbar}{4\rho^{1/2}\kappa^{1/2}}
 \left(\frac{1}{2\mu+\lambda}-\frac{1}{\mu}\right)
 \nonumber \\
 \int \frac{d\omega\ d^2 k}{(2\pi)^3}
 k_\alpha k_\mu k_\kappa k_\lambda
 h(\omega,\bm k) h(-\omega,-\bm k)
 \nonumber \\
 \int \frac{d^2 q}{(2\pi)^2 q^4}
 \left[\frac{\delta_{\beta\kappa} \delta_{\nu\lambda}}{q^2}
 -2 \frac{\delta_{\nu\kappa} q_\beta q_\lambda}{q^4} \right.
 \nonumber \\ \left.
 -2 \frac{\delta_{\beta\kappa} q_\nu q_\lambda}{q^4}
 -\frac{q_\beta q_\nu\delta_{\kappa\lambda}}{q^4}
 +4 \frac{q_\beta q_\nu q_\kappa q_\lambda}{q^6}\right]
 \nonumber \\
 (2\mu q_\alpha q_{\beta}
 +\lambda q^2 \delta_{\alpha\beta})
 (2\mu q_\mu q_{\nu}
 +\lambda q^2 \delta_{\mu \nu})
 \nonumber \\
 =\frac{\hbar}{16\pi\rho^{1/2}\kappa^{1/2}}
 \frac{(\mu+\lambda)(\mu+2\lambda)}{2\mu+\lambda}
 \ln\frac{\Lambda}{\Lambda'}
 \nonumber \\
 \int \frac{d\omega\ d^2 k}{(2\pi)^3} k^4
 h(\omega,\bm k) h(-\omega,-\bm k).
 \nonumber
 \end{eqnarray}
The ``isotropic'' contribution to the slow action is written as
 \begin{eqnarray}
 \Delta_{22}I^{(2)}= \frac{\hbar}{4\rho^{1/2}\kappa^{1/2}}
 \nonumber \\
 \int \frac{d\omega\ d^2 k}{(2\pi)^3}
 k_\alpha k_\mu h(\omega,\bm k) h(-\omega,-\bm k)
 \nonumber \\
 \int \frac{d^2 q}{(2\pi)^2 q^2}
 \left[\frac{k_\beta k_\nu}{q^2}
 -2 \frac{q_\beta k_\nu}{q^2}
 \frac{\bm q \bm k}{q^2} \right.
 \nonumber \\ \left.
 -2 \frac{k_\beta q_\nu}{q^2}
 \frac{\bm q \bm k}{q^2}
 +4 \frac{q_\beta q_\nu}{q^2}
 \frac{(\bm q \bm k)^2}{q^4}
 -\frac{q_\beta q_\nu}{q^2}
 \frac{k^2}{q^2}\right]
 \nonumber \\
 \left[\mu(q_\alpha q_\mu \delta_{\beta \nu}
 +q_\beta q_\mu \delta_{\alpha \nu}
 +q_\alpha q_\nu \delta_{\beta \mu}
 +q_\beta q_\nu \delta_{\alpha \mu}) \right.
 \nonumber \\ \left.
 +2\lambda (q_\alpha q_\beta \delta_{\mu \nu}
 +q_\mu q_\nu \delta_{\alpha\beta})
 +\frac{\lambda^2}{\mu}
 \delta_{\alpha\beta} \delta_{\mu \nu} \right]
 \nonumber \\
 =\frac{\hbar}{16\pi\rho^{1/2}\kappa^{1/2}}
 \left(-2\mu-2\lambda\right)
 \ln\frac{\Lambda}{\Lambda'}
 \nonumber \\
 \int \frac{d\omega\ d^2 k}{(2\pi)^3} k^4
 h(\omega,\bm k) h(-\omega,-\bm k).
 \nonumber
 \end{eqnarray}
Summing up the ``longitudinal'' and the ``isotropic'' contributions we find the same
correction to $\kappa$ as in real space calculations, leading to the RG-equation
(\ref{corr14}).

\end{document}